\documentclass[aps,prl,reprint,showpacs,floatfix,superscriptaddress]{revtex4-2}

\usepackage{amsmath,amsthm,amssymb}
\usepackage{graphicx}
\usepackage{dcolumn}
\usepackage{bm}
\usepackage{color}
\usepackage{epsfig}
\usepackage{multirow}
\usepackage{mathrsfs}
\usepackage{hyperref}
\usepackage{cleveref}
\usepackage{epstopdf}
\usepackage{subfigure}
\usepackage{autobreak}
\usepackage{orcidlink}

\begin{document}
\preprint{Preprint}

\title{Geometric and Dynamic Properties of Entangled Polymer Chains in Athermal Solvents: A Coarse-Grained Molecular Dynamics Study}

\author{Jiayi Wang}
\email{jwangfw@connect.ust.hk}
\affiliation{Thrust of Advanced Materials, The Hong Kong University of Science and Technology (Guangzhou), Guangdong, China}

\author{Ping Gao ~\orcidlink{0000-0003-0625-2391}} \thanks{Corresponding author}
\email{kepgao@ust.hk}
\affiliation{Department of Chemical and Biological Engineering, The Hong Kong University of Science and Technology, Clear Water Bay, Hong Kong, China}

\date{\today}

\begin{abstract}

Abstract: We used a coarse-grained model to study the geometric and dynamic properties of flexible entangled polymer chains dissolved in explicit athermal solvents. Our simulations successfully reproduced the geometrical properties including the scaling relationships between mean-square end-to-end distance $<R_{ee}^2>$, chain entanglement lengths $N_{e}$ and concentration $\Phi$. Specifically, we find that $<R_{ee}^2>\sim N*\Phi^{-1/4}$,
$N_{e} = 30.01\Phi^{-5/4}+31.23$. Dynamically, our model confirmed the ratio of the dynamic critical entanglement $N_{c}$ and the geometric entanglement length $N_{e}$ is constant, with $N_{c}/N_{e} = 5\sim 6$. To account for the local swelling effect for chains confined in athermal solvents, we treated the chains using the concept of blobs where each blob occupies a volume $\Omega_{b}$, with length $g$. Direct MD simulations and scaling analysis showed that $g \sim \Phi^{-25/36}$, $\Omega_{b}\sim\Phi^{-5/4}$. Using these together with the concentration dependent packing length $p \sim \Phi^{-5/12}$, we obtained a modified the Lin-Noolandi ansatz for concentrated flexible polymer chains in athermal solvents: $G \sim \frac{\Phi}{\left(N_{e} / g\right) \Omega_{b}} \sim  \Phi^{-2.28}$. We demonstrate this modified ansatz agrees well with our coarse-grained numerical simulations.
\end{abstract}



\maketitle

\section{\romannumeral1. INTRODUCTION}

The dynamics of polymer systems are significantly affected by entanglement caused by topological constraints between chains \cite{brown1996entanglements,graessley1971linear,graessley2005entanglement,kong2021control,likhtman2014microscopic}. Several models have been proposed to describe the dynamic properties of entangled polymer melts \cite{de1990introduction,doi2013soft,fetters1999packing,ferry1980viscoelastic,hoy2020unified,dietz2022validation}. However, there is a need to improve our understanding of polymers dissolved in athermal solvents regarding geometric and dynamics aspects. This is because the entanglement length $N_{e}$ and the local environment of the polymer chains will change due to the swelling effect caused by the athermal solvents in contrast to the pure melt. These effects, including the evolution of the entanglements and the local swelling of polymer chains by athermal solvents, are not readily observable experimentally, necessitating numerical simulations.

Various theoretical models have been proposed to quantify polymer entanglements. Geometric analyses by de Gennes, Kavassalis, and Richard P. Wool resulted in the scaling expression $N_{e} \sim N_{c} \sim \Phi^{-5/4}$ \cite{de1979scaling,de1976dynamics,kavassalis1987new,kavassalis1988new,kavassalis1989entanglement,wool1993polymer,russell1993direct}, where $N_{e}$ denotes the geometric entanglement length, $N_{c}$ represents the dynamical critical length of entanglement, and $\Phi$ is the polymer concentration. In terms of dynamics, the Rouse model predicts that the relaxation time of the entangled system is proportional to the square of chain length when the chain length is lower than $N_{c}$ \cite{de1976dynamics2}. However, the Rouse model cannot be applied to solution systems due to the leakage assumption, which is not strictly satisfied. To address this issue, the Zimm model considers polymer chains and solvents as a whole \cite{zwanzig1974theoretical,zimm1956dynamics}, and their motion satisfies the Stokes-Einstein relation \cite{shearer2008fluid}. The expression between relaxation time and chain length is obtained as $\tau \sim N^{3v}$, where $v$ is 0.588 in athermal solvents solution and $N$ is the monomer number of the polymer chain. When the chain exceeds $N_{c}$, the long-lived entanglement effect occurs because of the topological constraints between chains. Edwards and de Gennes introduced the concept of a confining tube \cite{edwards1967statistical,de1971reptation}, where the motion of the polymer chain is analogous to a snake wriggling inside a narrow tube, and deduced $\tau \sim N^{3}$. The relaxation time acts as a bridge that relates the elastic response to the viscous response of entangled polymers. When $t<\tau$ , the polymer exhibits an elastic response to external forces, acting like a cross-linked network, while $t>\tau$, the entanglements slip, resulting a viscous response. Regarding the elastic response of flexible polymer solutions, the Lin-Noolandi ansatz provides the relationship between the plateau modulus $G$ , $\Phi$ and $N_{e}$: $G \sim k_{B}T\Phi/(N_{e}\Omega_{0})$, where $k_{B}$ is the Boltzmann constant, $T$ refers to temperature, and $\Omega_{0}$ is the volume of a monomer \cite{lin1987number,milner2020unified}. This relationship has not been verified for flexible polymers in athermal solvents where excluded volume effect may affect the chain stiffness.

In this study, we investigated the geometric and dynamic properties of flexible polymer chains dissolved in the explicit athermal solvents using large-scale MD simulations.  To model the local swelling effects in such systems, we introduced the concept of swelling blobs and derived scaling relationships for the entanglement length $N_{e}$, the packing length $p$, the blob length $g$ and volume $\Omega_{b}$ by combining MD simulation with the blobs scaling concept. We modified the Lin-Noolandi ansatz to incorporate the swelling blob scaling relationships, resulting a new scaling relationship: $G \sim \frac{\Phi}{\left(N_{e} / g\right) \Omega_{b}}\sim \Phi^{2.28}$. The modified scaling relationship matches closely to that observed in experiments \cite{milner2020unified,graessley2005entanglement}.

\section{\romannumeral2. SIMULATION METHOD}
We modelled polymer chains using the bead-spring model \cite{liu2008effective}, which assumes that the polymer chain is composed of $N$ sequentially connected beads (monomers). To enhance the simulation efficiency for large systems, our model only considers the repulsive non-bonded energy between monomers and the bonding energy connecting two adjacent monomers only.  The chains are fully flexible so the the bending energy of two adjacent bonds is taken as zero. 
The nonbonded interaction between monomers is described by the purely repulsive Weeks-Chandler-Andersen (WCA) potential \cite{weeks1971role}:
\rightline{$U_{L J}(r)=\left\{\begin{array}{ll}
4 \varepsilon_{0}\left[\left(\frac{\sigma}{r}\right)^{12}-\left(\frac{\sigma}{r}\right)^{6}\right]+\varepsilon_{0} & r \leq 2^{1 / 6} \sigma \\
0 & r>2^{1 / 6} \sigma
\end{array}\right. \quad (1)$}
where $r$ is the distance between two monomers, $\varepsilon _{0}$ is the interaction strength, $\sigma$ is the diameter of monomer. At high temperatures, the presence of attractive interactions between monomers has little effect on the dynamical properties of the polymer melts. Therefore, the WCA potential saves a large amount of computational time by retaining only repulsive interactions and achieves high accuracy. The finitely extensible nonlinear elastic (FENE) potential is used to simulate the bonding energy \cite{kremer1990dynamics}:
\rightline{$U_{F E N E}(r)=-\left(k_{F E N E} R_{0}^{2} / 2\right) \ln \left[1-\left(\frac{r}{R_{0}}\right)^{2}\right] \quad\quad (2)$}
where $k_{FENE}$ refers to the spring constant, $r$ and $R_{0}$ represent the bond length and the maximum bond length, respectively. The chains were modelled using an algorithm for self-avoidance walk (SAW), and our algorithm was modified so that each chain not only avoided itself but also avoided other chains. 

The modeling process is given in Fig. 1(a)-(c), after the construction of the initial model by the SAW method, a soft-core potential was used to avoid the singularity in the energy minimization process, making the energy surface smoother. Following the treatment of the soft potential, the full force field was applied to the system for the final energy minimization and system relaxation.To describe the solution of fully flexible chains in athermal solvents, we modelled the solution by taking the analogy with the Flory lattice model, i.e., each solvent molecule is treated as a lattice point, and a polymer chain of length $N$ is treated as a connection of $N$ lattice points \cite{sariban1987critical,sayegh1980lattice}. In other words, we modelled the solutions to satisfy the following conditions:
 \rightline{$\varepsilon _{MM}=\varepsilon _{MS}=\varepsilon _{SS}=\varepsilon _{0} , \quad \sigma_{S}=\sigma _{M}=\sigma$   \quad (3)}
where $\varepsilon_{MM}$, $\varepsilon_{MS}$ and $\varepsilon_{SS}$ refer to the interaction of monomer-monomer, monomer-solvent and solvent-solvent respectively, $\sigma_{S}$ and $\sigma_{M}$ describe the diameter of solvent beads and chain monomers. When equation (3) holds, the polymer chains can be fully swollen by the athermal solvents, and the chains in such system are expected to follow the scaling rule suggested by de Gennes \cite{de1979scaling}.
\begin{figure*}[htbp]
\includegraphics[width = 1.0\textwidth]{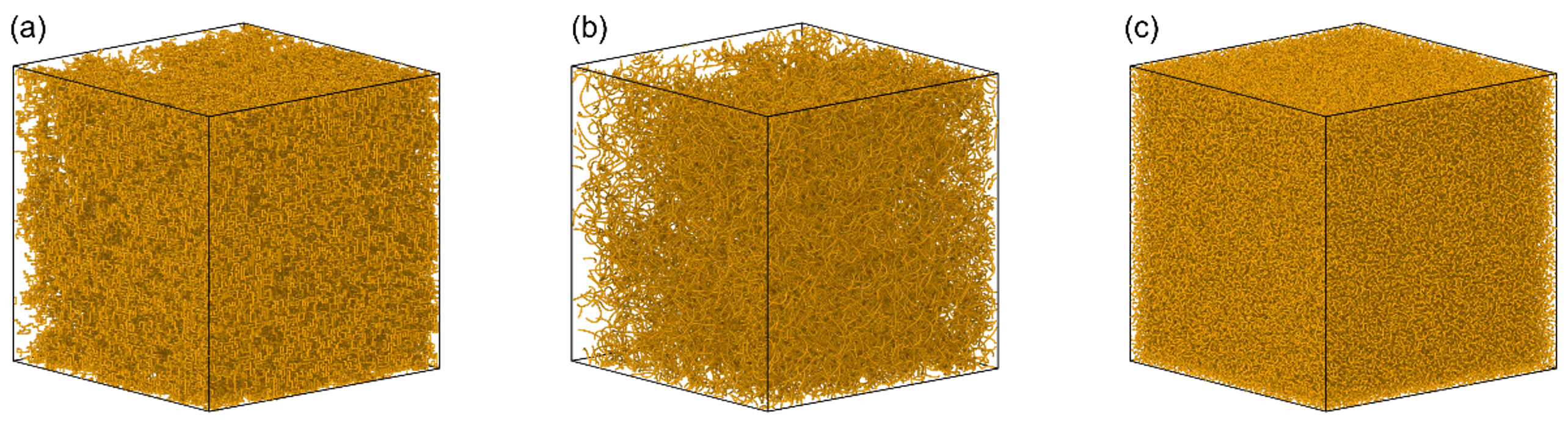}
\caption{Model construction. (a) Self-avoidance walk algorithm to build the initial model, 3-D periodical boundary condition is imposed. (b) A soft-core potential is used to facilitate the relaxation process. (c) Relaxation process under full force field.}
\label{FIG. 1 }
\end{figure*}

Here, $m_{monomer}=1 \quad g/mol$, $\varepsilon_{0}=1\quad kcal/mol$ and $\sigma=5 \quadÅ$. As regards the FENE potential, we set $k_{FENE} = 30 \quad \varepsilon_{0}/\sigma$ and $R_{0}=1.5\quad\sigma$ \cite{hoy2020unified}. Since the periodic boundary condition is introduced in three dimensions, large-scale models are required so that the size of the box is greater than the root mean end-to-end distance $<R_{ee}>$ of the polymer chain. With such a setup, the conformation of each chain will not be affected by the box boundaries, each of our models contains $10^{6}$ beads and the total number of beads exceeds $10^{7}$. Number density is set to $0.7/\sigma^{3}$ and temperature is $2.0\quad \varepsilon_{0}/k_{B}$, where $k_{B}$ is the Boltzmann constant. The Nosé-Hoover thermostat is used to control the temperature with a timestep of 1 $fs$ in the NVT ensemble \cite{evans1985nose,nauchitel1981energy}, with simulation time up to 500 $ns$ until chains are fully relaxed. All simulations were carried out using the Large-scale Atomic/Molecular Massively Parallel Simulator (LAMMPS) software \cite{plimpton1995fast}.
\section{\romannumeral3. RESULT AND DISCUSSION}
\subsection{\emph{A.Entanglement length and chain relaxation dynamics - Pure Melt System}}
Two length scales, a geometric entanglement length, $N_{e}$, and a dynamic critical entanglement length, $N_{c}$, are commonly used to describe entangled polymer melts \cite{ fetters1999packing}. The relationship between $N_{c}$ and $N_{e}$ has been studied by de Gennes, Kavassalis and Richard P. Wool, and a constant ratio between $N_{c}$ and  $N_{e}$ has been estimated to be between 9/4 to 27/4 for isotropic melt of flexible polymer chains \cite{de1979scaling,kavassalis1987new,wool1993polymer}. In this work, we hope to establish this constant ratio through molecular dynamics simulations and verify whether it holds for fully flexible chains in the pure melt and in athermal solvents. 

\begin{figure}[h]
\centering
\includegraphics[width = 0.45\textwidth]{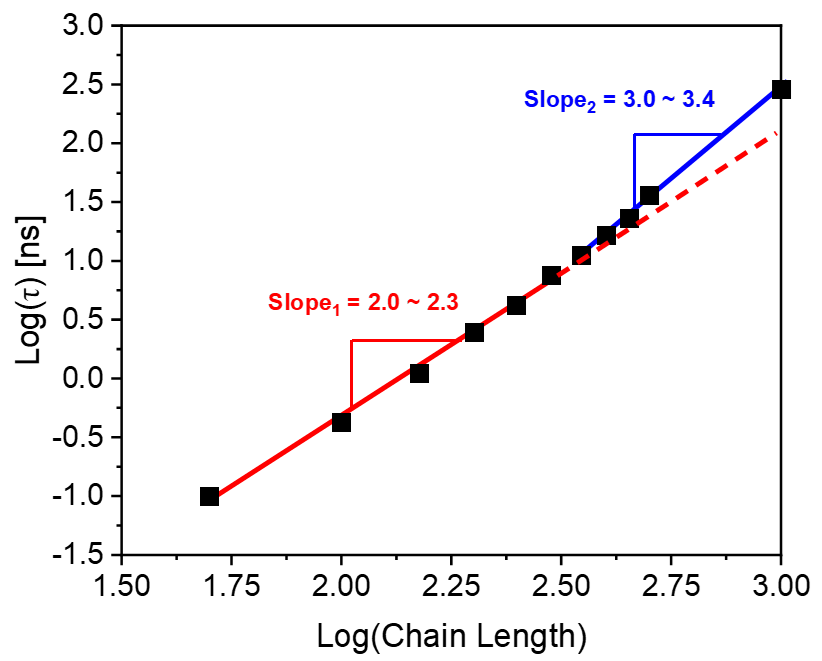}
\caption{The relationship between relaxation time and chain length, with the filled squares representing the simulated data, and the continuous lines showing the fitted lines depicting slopes of $2.0\sim2.3$ and $3.0\sim3.4$, respectively.  The relaxation time obeys the Rouse model of the unentangled melt for chains of lengths below the critical size marked at the beginning of the dashed red line. Above the critical chain length, the relaxation time coincides with the tube model of de Gennes and Doi and Edwards.}
\label{FIG. 2 }
\end{figure}
We generated models of monodisperse polymer melts with varying chain lengths, ranging from 100 to 1000 monomers per chain to cover the unentangled to entangled range. To determine the entanglement length, we used the Z1 method developed by Martin Kr{\"o}ger\cite{kroger2005shortest,hoy2009topological,shanbhag2007primitive}, which yielded a constant value of $N_{e}\cong58$, consistent with the previous literatures \cite{grest2016communication,cao2015simulating}. To calculate $N_{c}$, we computed the relaxation time for each model by analyzing the autocorrelation function of the end-to-end vector.  The resulting relaxation time is plotted in Fig. 2, with the red and blue lines representing the unentangled and entangled states, respectively.  Our analysis reveals a critical chain length $N_{c}$ above which the proportionality constant between the logarithmic relaxation time of the melt and that of the chain length changes from $2.0\sim2.3$ to $3.0\sim3.4$. The critical length $N_{c}$ was approximately 316 based on the figure, and the unentangled and entangled regions were consistent with the Rouse and tube models, respectively. Thus, the ratio of the critical dynamical entanglement length $N_{c}$ over that of the geometric entanglement length $N_{e}$  equals to 5.45. This result lies in the range of 9/4 to 27/4 as given by the geometric analysis of Wool \cite{wool1993polymer}.

\subsection{\emph{B.Entanglement length and chain relaxation \quad dynamics - Solution System}}
To ensure the accuracy of our simulation model for fully flexible polymer chains in athermal solvents, we first validated our coarse-grained model by simulating the scaling laws for the mean square end-to-end distance $<R_{ee}^{2}>$ and the entanglement length $N_{e}$ at varying polymer volume fractions $\Phi$. The scaling law for $<R_{ee}^{2}>$ at concentrations well above the over lapping concentration $\Phi^{*}$, proposed by de Gennes, is given by equation (4), as follows \cite{daoud1975solutions}: 
\rightline{$<R_{ee}^{2}> \sim N^{1.0}\Phi^{-0.25}$ \quad\quad\quad\quad\quad\quad(4)}

\begin{figure}[h]
\centering
\includegraphics[width = 0.49\textwidth]{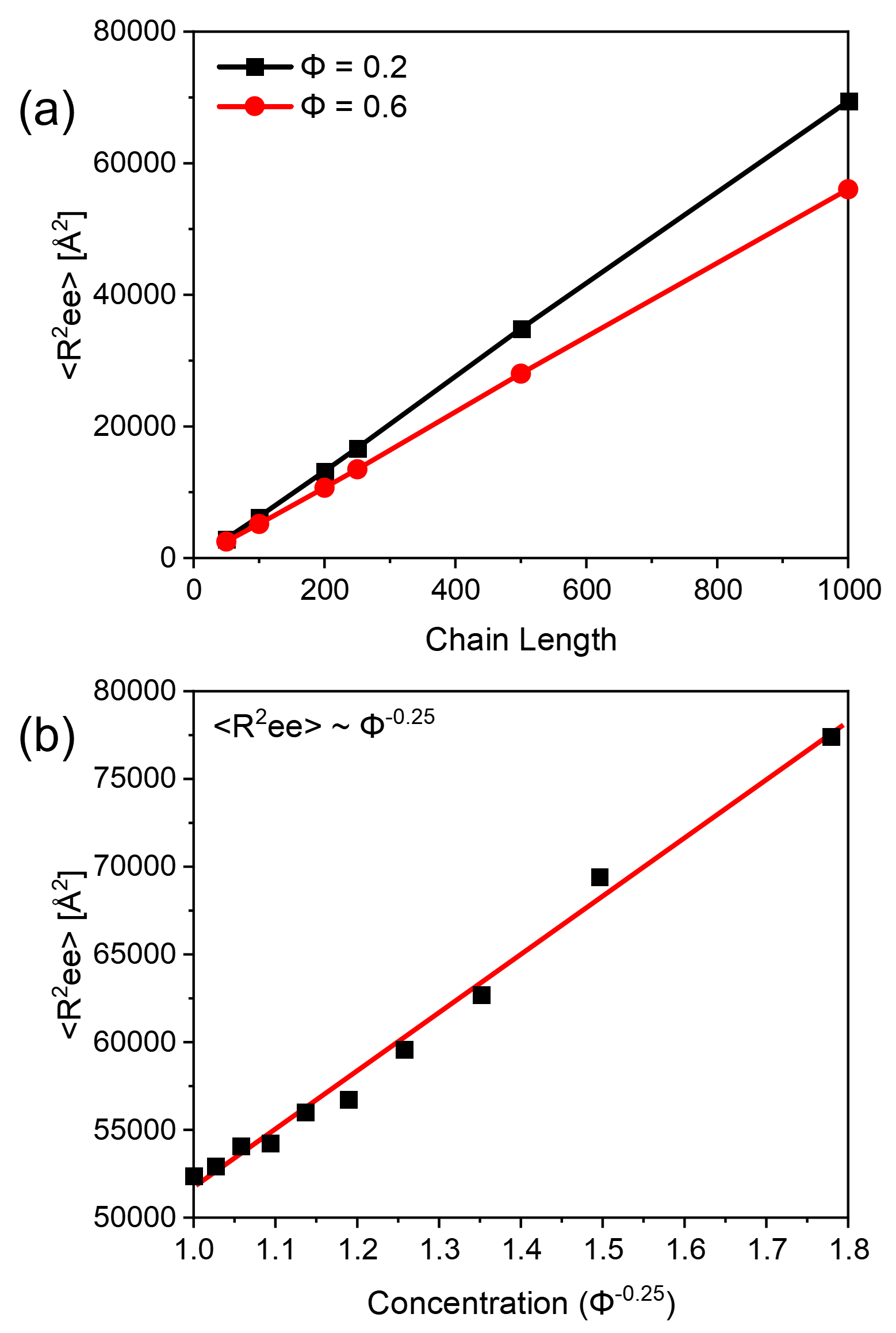}
\caption{Scaling relationships for the $<R_{ee}^{2}>$ versus chain lengths at fixed concentration (a) and polymer concentrations at fixed chain length of $N = 1000$(b). The concentrations of 0.2 and 0.6 chosen for (a) represent the values below and above the dynamical critical concentration of N = 1000 model. The data points in both (a) and (b) correspond to the simulated results, while the linear lines represent the prediction of equation (4).}
\label{FIG. 3 }
\end{figure}
Here $N$ represents the polymer chain length and $\Phi$ is the polymer volume fraction. We investigated the dependence of dependence of $<R_{ee}^{2}>$ on chain length and polymer concentration by MD simulations. Six models of monodisperse polymer chains were generated at $N=50,100,200,250,500$ and $1000$,with " fixed polymer concentrations at $\Phi$ = 0.2 and 0.6 for each model. Additionally, ten models of monodispersed polymer chains were generated at $N=1000$, but varying polymer volume fractions from 0.1 to 1.0 at constant intervals of 0.1 to explore concentration dependence. All simulation range corresponds to the heavily overlapped systems as they are well above the critical concentration for chain overlap, $\Phi^{*}_max=0.068$ when $N=50$. We plotted the calculated values of $<R_{ee}^{2}>$ as a function of $N$ and $\Phi^{-0.25}$, in Figures 3(a)-(b), respectively. The data points represent the simulation values, while the lines were obtained by linear fitting of equation (4). The excellent agreement between the calculated values to the fitted lines given by equation (4) shows that our model can successfully describe the concentrated polymer solutions in athermal solvents. 

\begin{figure}[h]
\centering
\includegraphics[width = 0.45\textwidth]{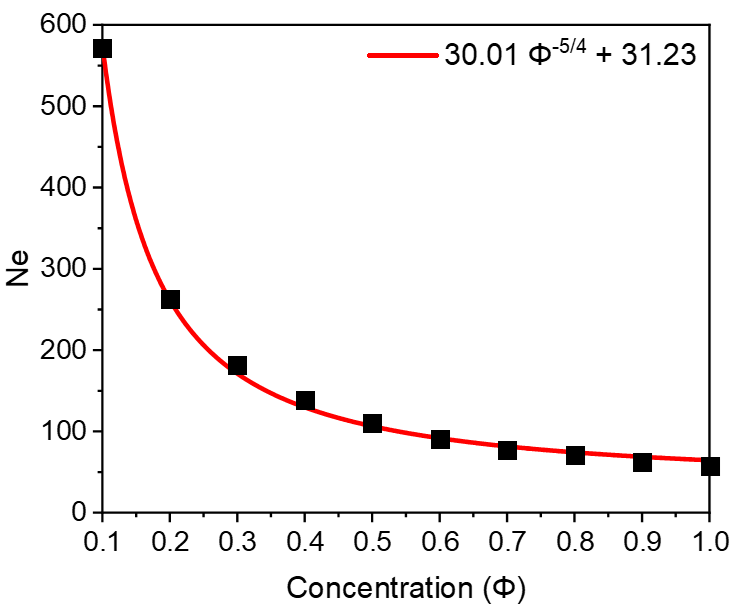}
\caption{The entanglement length of concentrated polymer chains in athermal solvents at different concentrations, calculated by the Z1 method.}
\label{FIG. 4}
\end{figure}

For the entanglement length, using the Z1 method, we calculated the entanglement length $N_{e}$ of $N=1000$ models at different concentrations and plotted the results in Fig. 4. The data points represent the simulated results, and the continuous line represents the fitting of $N_{e}=A \Phi^{-1.25}+B$, where $A$ and $B$ are fitting constants. Whilst the scaling exponent agrees with the literature findings of $N_{e} \sim \Phi^{-5/4}$ for flexible polymer chains in good solvent systems \cite{de1979scaling}, a non-zero constant $B$ was observed in this study.  This is because that polymer chains are unentangled at concentrations below the overlapping concentration $\Phi^{*}$.  

After demonstrating the accuracy of our coarse-grained models in predicting the  geometric scaling laws for concentrated polymer solutions in athermal solvents, we proceeded to simulate the dynamic relaxation behavior of these polymer solutions of monodisperse polymer chains at $N = 1000$. We computed the relaxation time constant by analyzing the auto-correlation function of the end-to-end vector. Fig. 5 (a) plots the simulated logarithmic relaxation time versus logarithmic polymer concentrations, with the scattered points representing the simulations and the continuous lines fitted by least square linear fitting.  The best fit to the data shows a slope change at a critical concentration $\Phi_{c}=0.3$. 

From the scaling relationship established in Fig. 4, we can deduce this corresponds to $N_{e}=182$.  Assuming this corresponds to the long-lived entanglement effect, we may deduce the dynamic critical entanglement length $N_{c}=1000$, giving a ratio of $N_{c}/N_{e} =5.49 = 5\sim6$.  This ratio is nearly the same as that simulated for the pure melt system presented in the previous section. Therefore, we may conclude that both the pure melt and fully flexible polymer chains in athermal solvents show similar ratios between geometric and dynamic entanglement lengths. 

\begin{figure}[t]
\centering
\includegraphics[width = 0.48\textwidth]{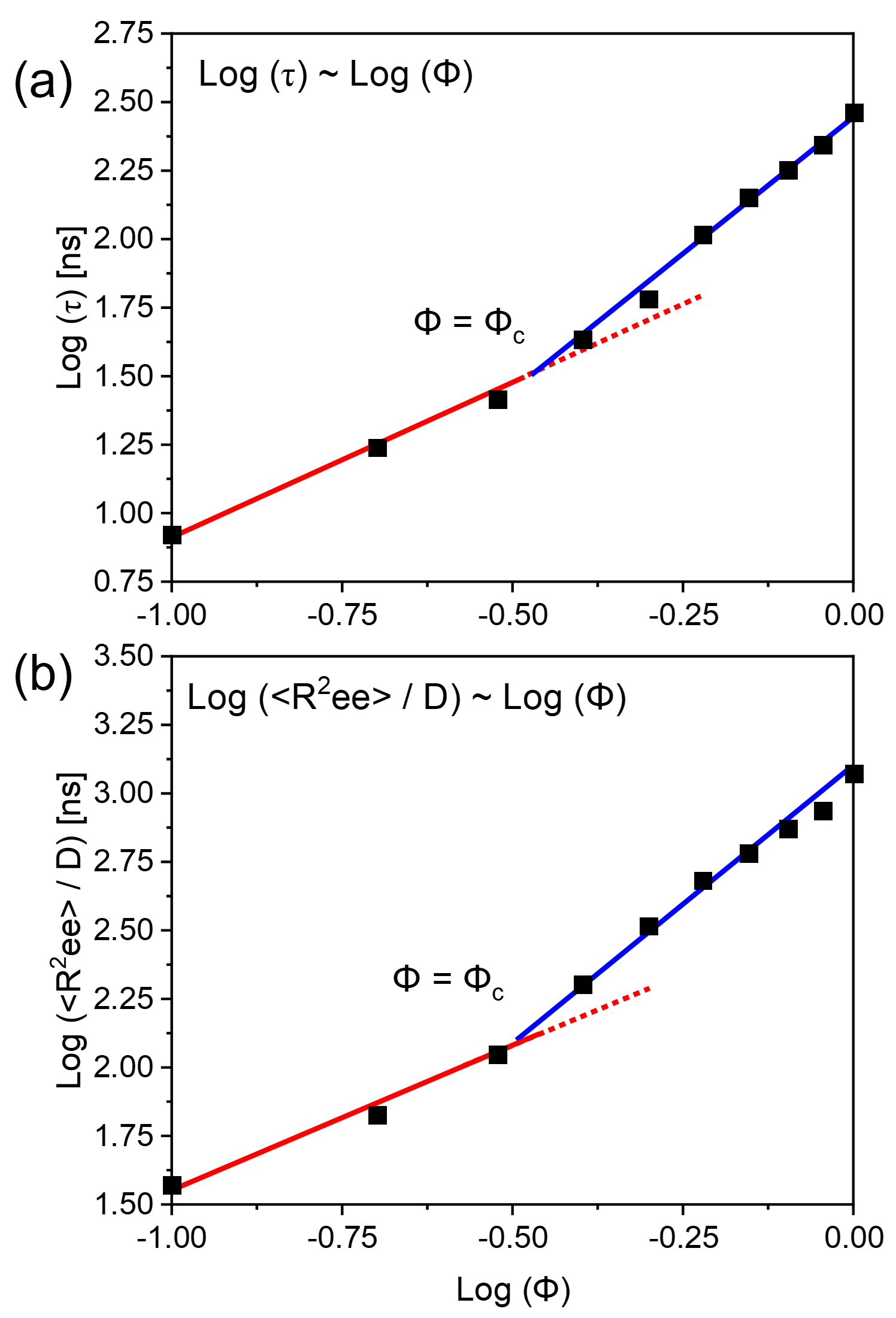}
\caption{Logarithmic relationships between relaxation time and polymer chain concentration in athermal solvents. (a) Simulation of relaxation time by auto-correlation of end-to-end chain vector; and (b) Relaxation time computed from diffusion dynamics simulation; the data points are simulated from our coarse-grain modelling for concentrations from 0.1 to 1.0, while the solid lines correspond to linear fitting.}
\label{FIG. 5}
\end{figure}

To further validate our analysis, we computed the relaxation time constant of the polymer solution by computing the diffusion time constant. The relaxation time corresponds to the time required for a polymer chain displacement over a length scale equals to that of the radius of gyration or mean end-to-end distance.  The simulated results are plotted in Fig. 5(b).  Like Fig. 5(a), the fitted lines also show a slope change at a critical concentration $\Phi_{c}=0.3$, and the values of the two slopes are identical to those in Fig. 5(a).  This further shows the accuracy and the consistency of our simulation results.

Figs. 6(a)-(b) illustrate the scaling relationships between diffusion coefficient and polymer concentration below and above the critical concentration for polymer chains in athermal solvents, respectively.  These plots demonstrate excellent agreement between the calculated results and the literature scaling relations for unentangled and entangled polymer solutions \cite{de1990introduction,doi2013soft,de1979scaling}, highlighting the ability of our model to accurately simulate the dynamics of polymer solutions. Our simulations of dynamic responses, in terms of scaling relations of diffusivity and relaxation time constant, have uncovered a new critical concentration $\Phi_{c}$, above which a long-lived entanglement effect becomes significant. Importantly, this critical concentration is significantly larger than the critical concentration for chain overlap. We propose that $\Phi_{c}$ should be regarded as a new parameter to quantify the dynamical state of entangled polymers in athermal solvents.
 
\begin{figure}[t]
\centering
\includegraphics[width = 0.475\textwidth]{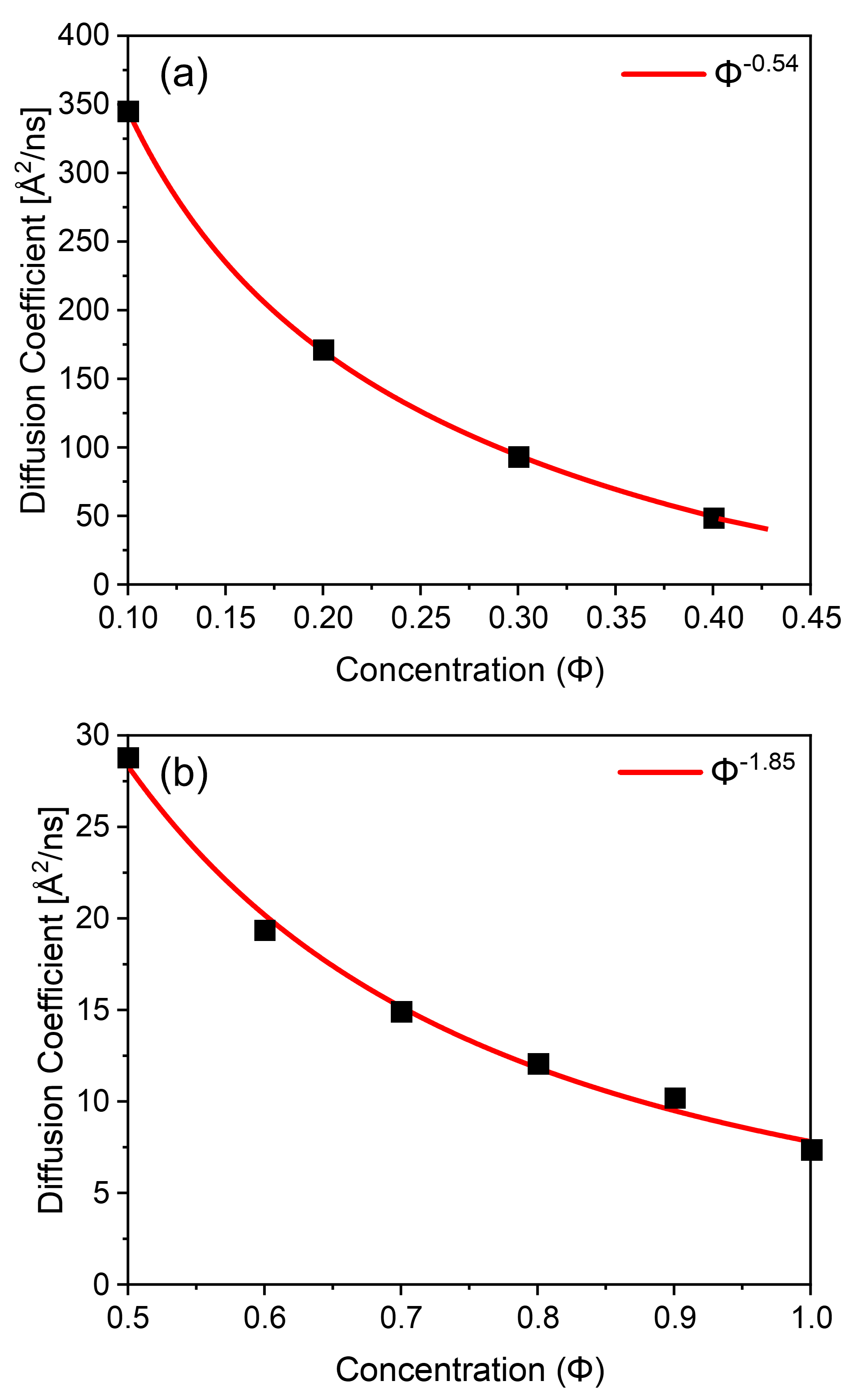}
\caption{Relationships between diffusion coefficient of flexible polymer chains in athermal solvents at concentrations below and above the critical concentration. (a) Below $\Phi_{c}$ with the continuous line represents the fitted data with $D\sim\Phi^{-0.54}$.  (b) Above $\Phi_{c}$ with the continuous line , $D\sim\Phi^{-1.85}$.}
\label{FIG. 6}
\end{figure}
\subsection{\emph{C.Local Swelling Effect}}

Polymer chains in athermal solvents are strongly influenced both by the solvents and adjacent chains. The size of a target polymer chain becomes swelled by the solvents and follows the law of self-avoidance walk (SAW) in dilute solutions.  In concentrated polymer solutions, the screening effect of proximity chains causes the chains to exhibit the Gaussian scaling in the whole. Such screening effect implicates that the swelling effects are limited to the local scale. The local environment of a chain can be described by packing length, which is defined by Milner as the closest distance between to backbone moieties on different strands. A larger packing length means there are more solvent molecules and fewer adjacent polymer chains around a target polymer chain. Milner’s definition enables the packing length to be readily evaluated from the radial distribution function (RDF), and accurate values of the packing lengths $p$ for flexible polymers of different architectures were obtained accordingly \cite{milner2005predicting,bobbili2021measuring}. The scaling relations between the packing length and polymer concentration can be derived following the analysis by Y.H. Lin, who stipulates that for all flexible polymers the number of entanglement strands coexisting in the volume penetrated by one entangled strands remains constant \cite{lin1987number}, i.e., 

\rightline{$p^{3} \sim N_{e}$ \quad\quad\quad\quad\quad\quad\quad\quad\quad  (5)}
\begin{figure}[t]
\centering
\includegraphics[width = 0.48\textwidth]{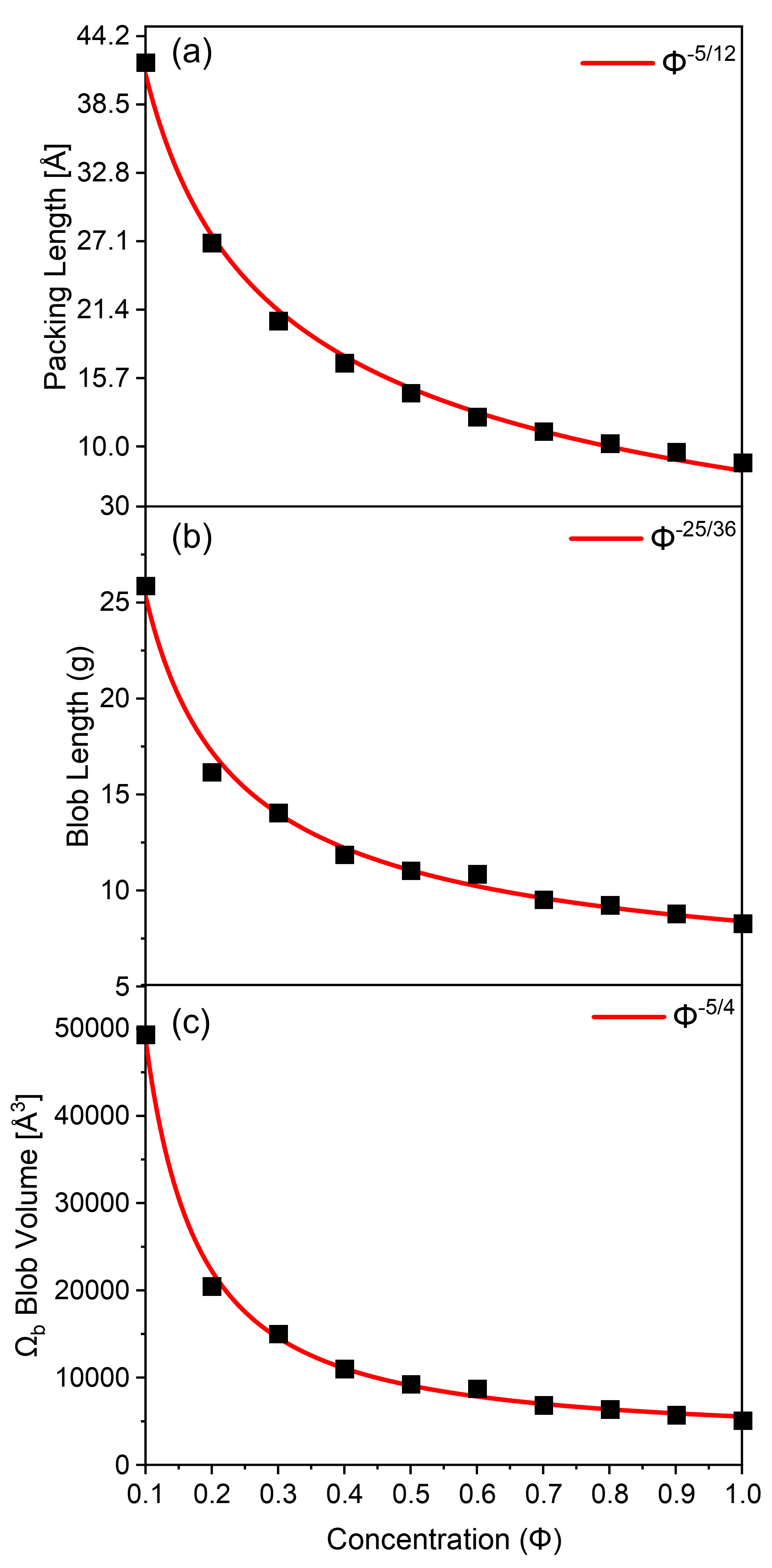}
\caption{Scaling relations for packing length $p$ (a), blob length $g$ (b) and blob volume $\Omega_{b}$ (c) versus concentration $\Phi$. The fitted lines are predicted by equations (6), (9) and (10), while the data points are from our coarse-grained modelling.}
\label{FIG. 7 }
\end{figure}
Here $p$ is the packing length, and $N_{e}$ is the entanglement length. By substituting $N_e \sim \Phi^{-5/4}$ to equation (5), the packing length is shown to scale with the polymer concentration as follows: 
     \rightline{$p \sim \Phi^{-5/12}$  \quad\quad\quad\quad\quad\quad\quad\quad     (6)}

The packing length of polymer chains at different polymer concentrations was calculated using the radial distribution function (RDF) method proposed by Milner, and the result was plotted together with that predicted using equation (6) in Fig. 7(a). The agreement between the simulation and the scaling prediction is excellent. 

To capture the local swelling effect in our model, we proposed a hypothesis that the polymer chain can be viewed as a series of connected swelling blobs, each with a diameter of $\xi$ and a length of $g$. The length of each blob, $g$, can be described by the correlation function of bond vectors. A larger $g$ corresponds to a stiffer chain, as it indicates a stronger correlation between bond vectors. We assumed that the correlation between chain segments in different blobs is lost due to the screening effect, while within a blob, the chain segment cannot effectively perceive the presence of other chains, resulting in swelling by the athermal solvents.
By assuming the blob size scales linearly with the packing size $p$, we find:

\rightline{$\xi \sim \Phi^{-5/12}$      \quad\quad\quad\quad\quad\quad\quad\quad   (7)}
And the coil diameter scales with the correlation length $g$ by:

\rightline{$\xi \sim g^{3/5}$  \quad\quad\quad\quad\quad\quad\quad\quad\quad(8)}

Combine equations (7) and (8), we arrived at the concentration scaling expression for the length $g$ as follows:

\rightline{$g\sim\Phi^{-25/36}$\quad\quad\quad\quad\quad\quad\quad\quad(9)}

Thus, the blob volume $\Omega_{b}$ is: 

\rightline{$\Omega_{b} \sim \xi^{3}\sim \Phi^{-5/4}$\quad\quad\quad\quad\quad\quad (10)}

We plotted the scaling equations for $g$ and $\Omega_{b}$ together with the simulated data in Figs. 7(b)-(c), and observed excellent agreement.

\begin{figure}[h]
\centering
\includegraphics[width = 0.25\textwidth]{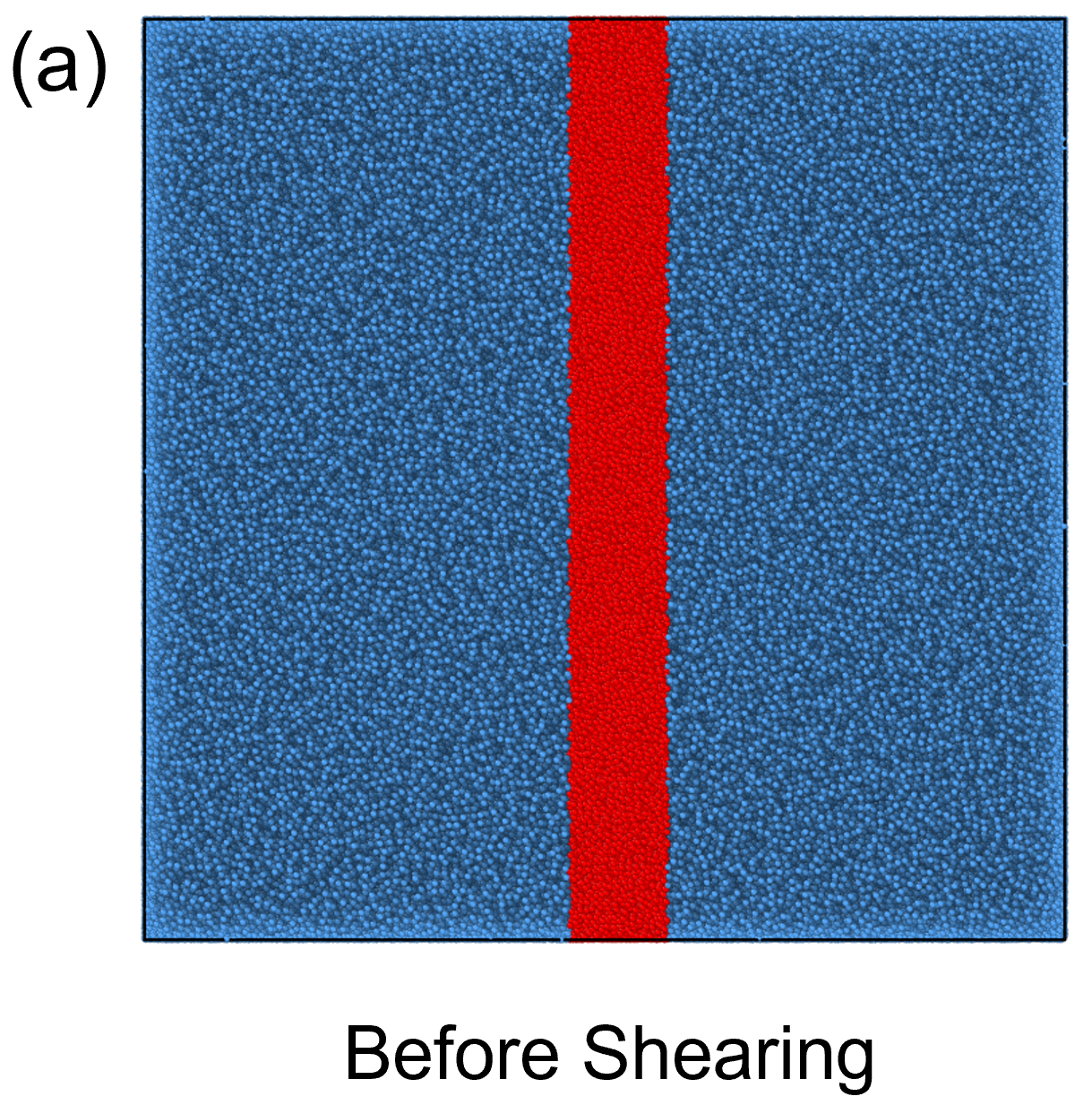}
\includegraphics[width = 0.48\textwidth]{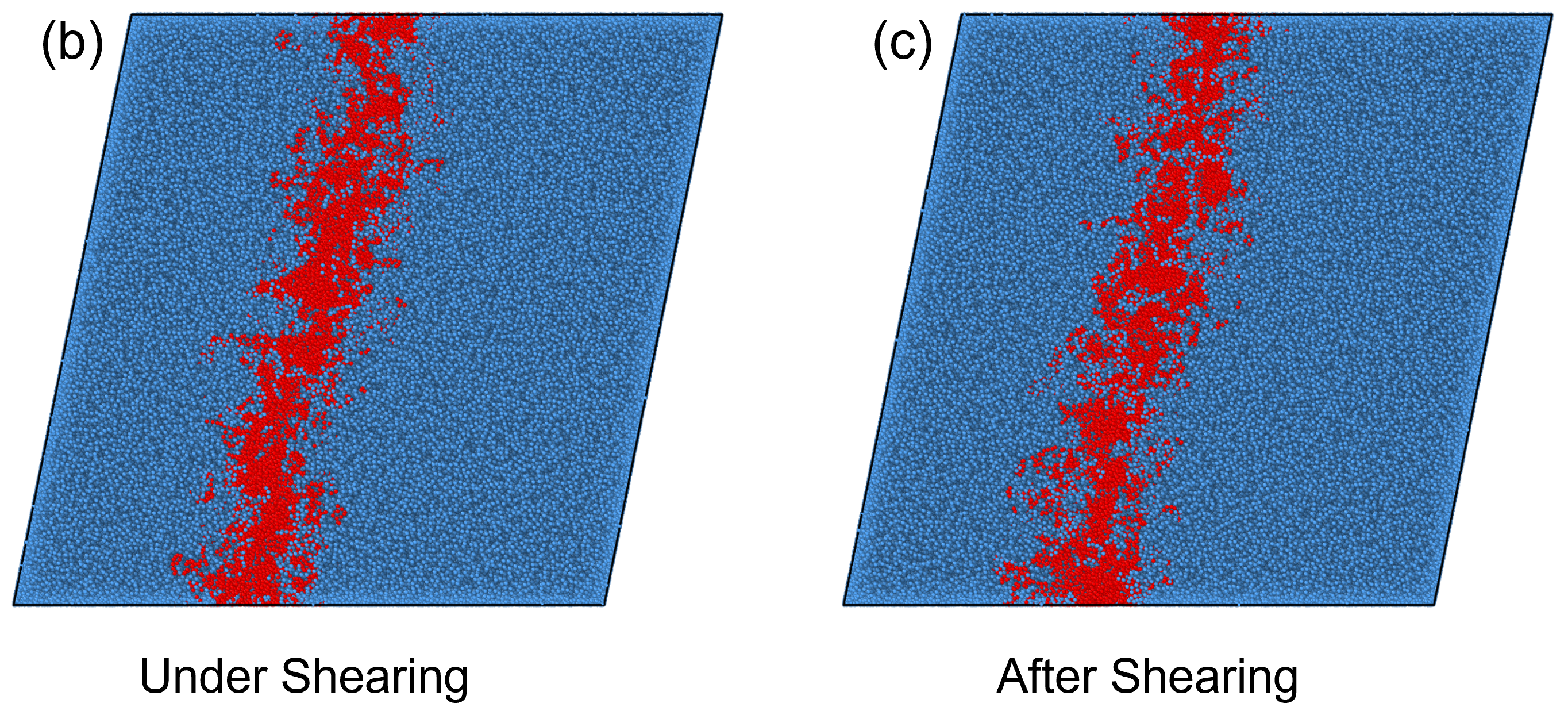}
\caption{Simulation of velocity profiles. (a)-(c) The system before, under and after shear flow. The middle section was highlighted in red to signify the velocity profile.}
\label{FIG. 8}
\end{figure}
\begin{figure}[h]
\centering
\includegraphics[width = 0.48\textwidth]{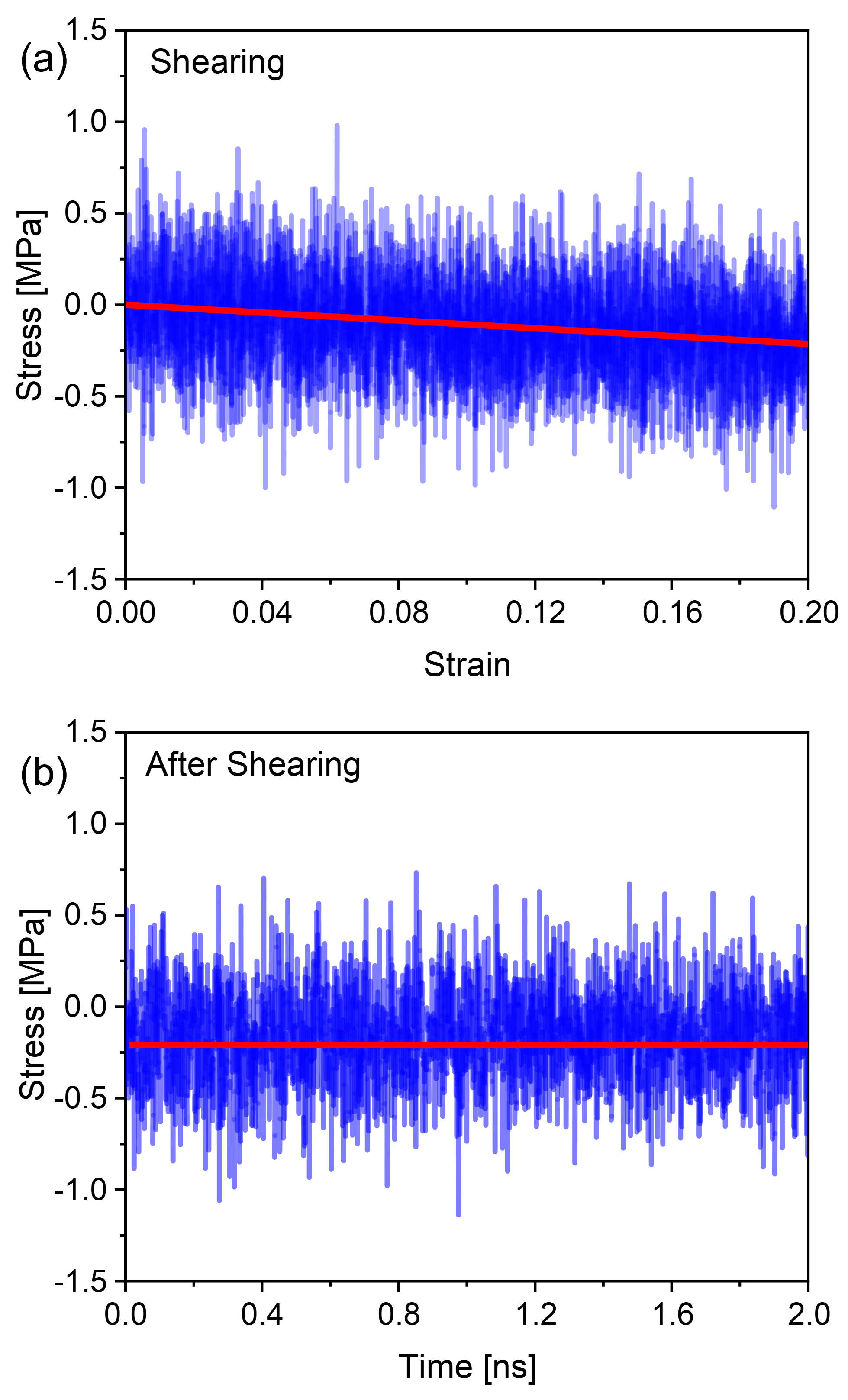}
\caption{(a) Stress-strain relationship in shearing process. (b) Stress-time relationship in relaxation progress.}
\label{FIG. 9}
\end{figure}

\begin{figure}[h]
\centering
\includegraphics[width = 0.48\textwidth]{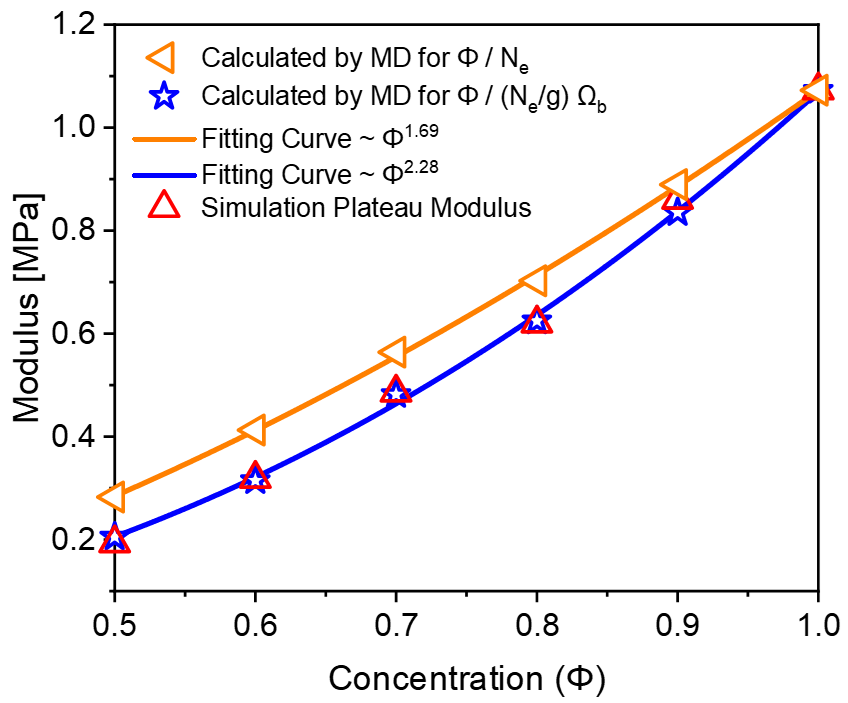}
\caption{
Comparison of plateau modulus $G$ versus concentration $\Phi$ between the fitting curve of modified Lin-Noolandi expression (blue line), the modified Lin-Noolandi Ansatz (blue stars), coarse-grain modelling (red triangles), and the classical Lin-Noolandi Ansatz ( orange line and triangle symbols).}
\label{FIG. 10}
\end{figure}

Our hypothesis, based on the swelling concept introduced previously, was that the elastic properties of the polymer chains could be altered, such that the plateau modulus of polymer chains $G$ in athermal solvents may not follow that by the classical Lin-Noolandi ansatz: $G \sim \Phi/N_{e}$. We have demonstrated that swelling of the chains inside the swelling blobs are stiffer as they exhibited larger correlation function $g$. In the meantime, the local swelling also leads to the formation of larger blob volumes, which render the chains softer. By taking these two factors into account, we obtained a modified scaling equation for the plateau modulus as follows:

\rightline{$G \sim \frac{\Phi}{\left(N_{e} / g\right) \Omega_{b}} \sim \frac{\Phi g}{N_{e} \Omega_{b}}$ \quad\quad\quad\quad\quad\quad{(11)}}

In addition, as described earlier (see Fig. 4), the scaling relation for $N_{e}$ should account for the non-entangling condition for dilute solutions, i.e., $N_{e} = A\Phi^{-1.25}+B$. Combining these together, we arrived at a new scaling expression for $G$ as $G \sim \Phi^{2.28}$. This expression agrees well with the previously reported experimental findings for concentrated polymer solutions in good solvent systems \cite{kavassalis1988new,kavassalis1989entanglement,graessley2005entanglement}. On the other hand, substitution of $N_{e} = A\Phi^{-5/4}+B$ into the classical Lin-Noolandi ansatz, we found that the scaling relation for $G$ as $G \sim \Phi^{1.69}$. This unexpected scaling exponent differs from previous literature reports \cite{milner2020unified}, we attribute the discrepancy to the non-negligible $B$, which had been previously ignored.

To investigate this numerically, we simulated the plateau moduli as a function of polymer concentration under simple shear deformation. Initially, we constructed a large-scale model box that was larger than the $<R_{ee}>$ of the chains, as shown in Fig. 8(a), to ensure the conformation of the polymers was not restricted by size. To simulate the shear progress, we deformed the box under a stable velocity along the $X$-direction\cite{cheal2018rheology,fang2018revealing,li2023stress,bobbili2020simulation}. Figs. 8(b) shows the stress-strain relationship in shear process, for a better observation, the beads in the middle of the system were colored in red. It can be observed that the system exhibited a uniform velocity field, the maximum shear strain was set to 0.2. After the shearing deformation, the system was relaxed at the fixed maximum strain, seen in Fig. 8(c), the chains in the system did not recover during the relaxation process. To further ensure the deformation was elastic, the stress evolution of the system during and after shearing are plotted in Fig. 9(a)-(b). From Fig. 9(a), the stress is linearly dependent on strain in the process of shearing, while in Fig. 9(b), when the system is fixed at the maximum strain, no stress relaxation is observed. These results suggest that the system was elastically deformed under the shear deformation. We computed the plateau modulus of the polymer chains at concentrations where the long-lived entanglements are important, i.e., for [$\Phi$ = 0.5, 0.6, 0.7, 0.8, 0.9, 1.0], and plotted the data in Fig. 10 along with the classical Lin-Noolandi ansatz and equation (11).

The orange and blue data points respectively represent the modulus calculated from the MD simulation of classical Lin-Noolandi ansatz and equation (11), while the red triangles represent the simulated modulus values from the shear simulation. Orange and blue curves show the fitting results of $G \sim \Phi/N_{e}$ and equation (11), resulting in $G \sim \Phi^{1.69}$ and $G \sim \Phi^{2.28}$, respectively. Our results show the simulated values deviate from the classical Lin-Noolandi ansatz but matches well with equation (11).  Both equation (11) and direct numerical modelling show excellent agreement with the concentration scaling law $G \sim \Phi^{2.28}$.  This new scaling relationship matches well with the experimental results $G \sim \Phi^{2.0\sim2.3}$ \cite{kavassalis1987new,kavassalis1988new,kavassalis1989entanglement,milner2020unified,graessley2005entanglement}, suggesting that large-scale MD simulations are important to study the underlying mechanisms for entangled polymer systems.

\section{\romannumeral4. DISCUSSION}

Our study demonstrates that our coarse-grained model can accurately describe the geometric scaling relationships for concentrated solutions of fully flexible polymer chains in athermal solvents. Through large-scale molecular dynamics simulations, we were able to verify and extend theories proposed by de Gennes, Kavasslis, Milner, etc. We also investigated the dynamic behavior of both pure melts and solutions, finding a consistent relationship between dynamic and geometric entanglement length. We identified a critical concentration above the crossover concentration which allows the solution to exhibit prolonged entanglement effects. Furthermore, a local swelling effect was discovered, using the concept of blob, we successfully correlated this local swelling with entanglement length $N_{e}$, packing length $p$, blob length $g$ and volume $\Omega_{b}$, and deduced the scaling relationships with concentration. As a result, we modified the Lin-Noolandi ansatz due to this local swelling effect.

We acknowledge that our coarse-grained model does not consider certain structural details of chains and solvents, and the drag effect and swelling mechanism remain unresolved issues. However, these limitations do not appear to affect our modelling capability, and further studies are needed to address these issues. Additionally, in this work, we focused only on the properties of flexible chains. We predict that there will be different dynamics in semi-flexible and stiff chains, and we plan to explore this issue in future work.
\bibliography{paper}
\end{document}